\documentclass[showpacs,amsmath,amssymb,aps,prl,twocolumn]{revtex4-1}

\usepackage{graphicx}
\usepackage{dcolumn}
\usepackage{bm}

\begin{document}

\title{Strain engineering magnetic frustration in perovskite oxide
  thin films}

\author{Carlos Escorihuela-Sayalero, Oswaldo Di\'eguez, and Jorge
  \'I\~niguez}

\affiliation{Institut de Ci\`encia de Materials de Barcelona
  (ICMAB-CSIC), Campus UAB, 08193 Bellaterra, Spain}

\begin{abstract}
Our first-principles results show that geometric frustration can be
induced in thin films of multiferroic BiFeO$_3$. We find that
competing magnetic interactions occur in the so-called {\em
  super-tetragonal} phase of this material, which can be grown on
strongly-compressive substrates. We show that the frustration level
can be {\em tuned} by appropriately choosing the substrate; in fact,
the three-dimensional spin order gets totally annihilated in a narrow
range of epitaxial strains. We argue that the effects revealed here
are not exclusive to BiFeO$_3$, and predict that they also occur in
multiferroic BiCoO$_3$.
\end{abstract}







\maketitle

The competition between different interactions underlies some of the
most fascinating phenomena in condensed-matter physics. When the
competing orders are of similar strength, rich phase diagrams are
likely to emerge, and the materials tend to be strongly responsive to
external perturbations. Examples abound in the family of perovskite
oxides, ranging from magnetoresistive manganites~\cite{tokura99} to
ferroelectric relaxors~\cite{pirc99,samara03}. In addition to the
fascinating science they involve, competing interactions often lead to
important functionalities.

Of special interest are the cases in which the competition relies on
the topology of the crystal lattice, as in a triangular network of
anti-ferromagnetically coupled spins, which is the classic example of
{\em geometric frustration}. Here we present first-principles results
showing that {\em strain engineering} (i.e., taking advantage of the
mismatch stress exerted by a substrate on a thin film) can be used to
induce a novel kind of {\em tunable} geometric frustration in a spin
system. Our work focuses on room-temperature multiferroic BiFeO$_3$,
and we show that the same effects also occur in other compounds like
BiCoO$_{3}$.

{\sl Tunable frustration in BiFeO$_3$.--} Multiferroic BiFeO$_3$ (BFO)
is a perovskite oxide that presents ferroelectric ($T_{\rm
  C}\approx$~1100~K) and magnetic ($T_{\rm N}\approx$~760~K) orders at
ambient conditions~\cite{catalan09}. While the usual BFO phase is
rhombohedral ($R3c$ space group), it was recently discovered that
BFO's atomic structure can be drastically modified by growing thin
films on strongly compressive (001)-oriented
substrates~\cite{bea09}. The phases thus obtained are called {\em
  super-tetragonal} ($ST$), as their (pseudo-cubic) unit cell presents
a very large aspect ratio $c/a\approx$~1.25 (see Fig.~1). Such
$ST$-BFO phases display a number of appealing features and are
currently receiving a lot of attention; in particular, they undergo
both structural and magnetic-ordering transitions slightly above room
temperature ($T_{\rm r}$)~\cite{infante11,ko11}, which might lead to
improved functional properties.

\begin{figure}[t!]
 \includegraphics[width=0.95\columnwidth]{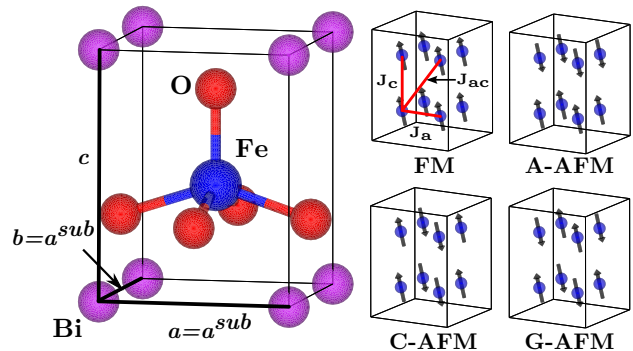}
 \caption{(Color online.) Left: Elemental 5-atom cell of $ST$-BFO,
   indicating the in-plane epitaxial constraint $a=b=a^{\rm
     sub}$. Right: Spin arrangements considered in this work. We
   sketch the 2$\times$2$\times$2 repetition of the elemental cell
   that we simulated, showing only the Fe atoms. In the FM case we
   indicate the exchange interactions $J$ discussed in the text.}
\end{figure}

The magnetic order of $ST$-BFO remains an open problem. Several
first-principles works~\cite{hatt10,dieguez11} predict the so-called
C-type anti-ferromagnetic (C-AFM) spin arrangement, while most
experimental studies suggest that the so-called G-AFM order
dominates~\cite{bea09,macdougall12}. As shown in Fig.~1, the C-AFM and
G-AFM orders are identical within the $ab$ plane (first-neighboring
spins are anti-parallel), but differ along the out-of-plane $z$
direction (first-neighboring spins are parallel in C-AFM and
anti-parallel in G-AFM). It is generally
accepted~\cite{hatt10,dieguez11,dieguez11b,infante11,macdougall12}
that the in-plane exchange interaction between neighboring Fe atoms
($J_a$ in Fig.~1) is anti-ferromagnetic and relatively strong; in
contrast, the out-of-plane couplings ($J_{c}$ and $J_{ac}$ in Fig.~1)
are believed to be small because of the very large separation between
irons along the $z$ direction. In fact, MacDougall {\sl et
  al}.~\cite{macdougall12} have argued that the occurrence of C-AFM or
G-AFM orders in specific films may be decided by factors that would be
{\sl extrinsic} to a perfect $ST$-BFO lattice, as the {\sl intrinsic}
out-of-plane couplings can be expected to be negligible.

\begin{figure}[t!]
\includegraphics[width=0.8\columnwidth]{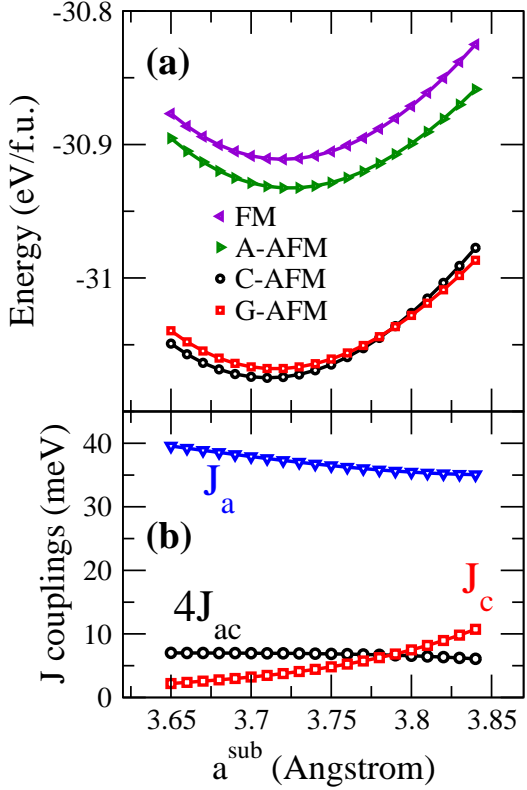}
 \caption{(Color online.) Panel~(a): Energies of the different spin
   arrangements as a function of $a^{\rm sub}$. Panel~(b): Exchange
   constants defined in Fig.~1 as a function of $a^{\rm sub}$. $J_{\rm
     ac}$ is multiplied by 4, to better visualize the point at which
   $E_{\rm G} = E_{\rm C}$.}
\end{figure}

Wanting to shed light on these issues, we used first-principles
methods~\cite{calcs,suppl} to investigate the magnetic order in
$ST$-BFO as a function of the epitaxial strain exerted by a
(001)-oriented square substrate. For this purpose, we considered a
perfectly tetragonal atomic structure ($P4mm$ space group); we checked
that, as regards the spin couplings, this structure is representative
of the variety of lower-symmetry (monoclinic)
phases~\cite{dieguez11,christen11} that occur in the actual films. For
each considered value of the substrate lattice parameter $a^{\rm
  sub}$~\cite{asub}, we studied several spin arrangements (FM,
A-AFM, C-AFM, and G-AFM, sketched in Fig.~1); for each arrangement, we
relaxed the atomic structure subject to the epitaxial constraint. As
shown in Fig.~2(a), we find that the C-AFM arrangement is the
lowest-energy solution for the most stable films, which correspond to
$a^{\rm sub}\approx$~3.71~\AA. The G-AFM order gets stabilized as
$a^{\rm sub}$ increases, the transition between C-AFM and G-AFM
occurring at $a^{\rm sub}\approx$~3.79~\AA.

To gain more insight, let us consider the Heisenberg spin Hamiltonian
$E - E_0 = 1/2N \sum_{i\neq j} J_{ij} \boldsymbol{S}_{i}\cdot
\boldsymbol{S}_{j}$, where $N$ is the total number of Fe atoms, needed
to capture such a crossover. Let us assume classical spins and take
$|\boldsymbol{S}_{i}|=1$ for simplicity. Then, if we restrict
ourselves to couplings between first-nearest neighbors (i.e., to
$J_{a}$ and $J_{c}$ in Fig.~1, as discussed by other
authors~\cite{hatt10,macdougall12}), the energies {\sl per} Fe atom of
the C-AFM and G-AFM orders are $E_{\rm C} = E_{0} -2J_{a} -J_{c}$ and
$E_{\rm G} = E_{0} - 2J_{a} +J_{c}$, respectively. The crossover point
($E_{\rm G} = E_{\rm C}$) would thus correspond to $J_{c} = 0$, i.e.,
to the value of $a^{\rm sub}$ at which $J_{c}$ changes sign. However,
this simple model is not satisfactory, for two main reasons: (1) It is
difficult to imagine an exchange mechanism that may render a
ferromagnetic (FM) coupling $J_{c}<0$. Indeed, the $J_{c}$ coupling is
associated to a Fe$^{3+}$--O$^{2-}$--Fe$^{3+}$ chain forming a
180$^{\circ}$ angle, for which the well-known Goodenough-Kanamori
rules~\cite{khomskii01} predict an AFM super-exchange interaction;
such an expectation should hold even in a case like this one, where
the two Fe--O bonds in the super-exchange path have different
lengths. (2) For $J_{c}<0$, such a simple model predicts $E_{\rm C} <
E_{\rm G} < E_{\rm F} < E_{\rm A}$. However, our results show that the
FM order is the least favorable one throughout the considered $a^{\rm
  sub}$ range. Indeed, our calculations predict that, even though
there is a crossover between the C-AFM and G-AFM orders, where the
magnetic interaction between $ab$ layers changes sign, the analogous
crossover between FM and A-AFM does not take place!

These difficulties are resolved by extending the model in the simplest
possible way, i.e., by including the $J_{ac}$ interaction defined in
Fig.~1. The energies {\sl per} Fe atom of our four magnetic orders are
then given by:
\begin{eqnarray}
E_{\rm F} & = & E_{0} + 2J_{a} + J_{c} + 4J_{ac} \, ,\\
E_{\rm A} & = & E_{0} + 2J_{a} - J_{c} - 4J_{ac} \, ,\\
E_{\rm C} & = & E_{0} - 2J_{a} + J_{c} - 4J_{ac} \, ,\\
E_{\rm G} & = & E_{0} - 2J_{a} - J_{c} + 4J_{ac} \, .
\end{eqnarray}
We can thus compute the $J$'s from the data in Fig.~2(a); the results
are shown in Fig.~2(b). This extended model is able to reproduce
exactly the first-principles energies of our four magnetic structures
in the whole $a^{\rm sub}$ range. In addition, the obtained values of
$J_{c}$ and $J_{ac}$ are always positive, which corresponds to AFM
interactions; as regards $J_{\rm c}$, this is compatible with the
expectations from the Goodenough-Kanamori analysis.

It may seem surprising that this model can capture the C-AFM ground
state, given that the computed out-of-plane interactions, $J_{c}$ and
$J_{ac}$, are both AFM in nature. To understand this, let us consider
first what determines the relative stability of the FM and A-AFM
orders, for which we have $E_{\rm F} - E_{\rm A} = 2J_{c} +
8J_{ac}$. Here, in both cases the $ab$ planes are ferromagnetically
ordered, and both $J_{c}$ and $J_{ac}$ favor the A-AFM solution over
the FM one. On the other hand, the energy gap between C-AFM and G-AFM
is $E_{\rm C} - E_{\rm G} = 2J_{c} -8J_{ac}$. In this case, the $ab$
planes are anti-ferromagnetically ordered, and the out-of-plane
interactions compete: A positive $J_{c}$ favors the G-AFM order {\sl
  via} the AFM coupling between Fe ions that are first neighbors
out-of-plane. In contrast, a positive $J_{ac}$ favors the C-AFM order
{\sl via} the AFM coupling between Fe ions that are second neighbors
out-of-plane. When $J_{ac}$ is large enough -- i.e., when $4J_{ac} >
J_{c}$, where the factor of 4 comes from the ratio between first- and
second-nearest neighbors out-of-plane--, the C-AFM order
prevails. Hence, according to this model, the predicted stabilization
of the C-AFM phase relies on {\em both} the AFM interaction $J_{ac}$
{\em and} the presence of a robust AFM order within the $ab$ planes.

\begin{figure*}[t!]
 \includegraphics[width=\textwidth]{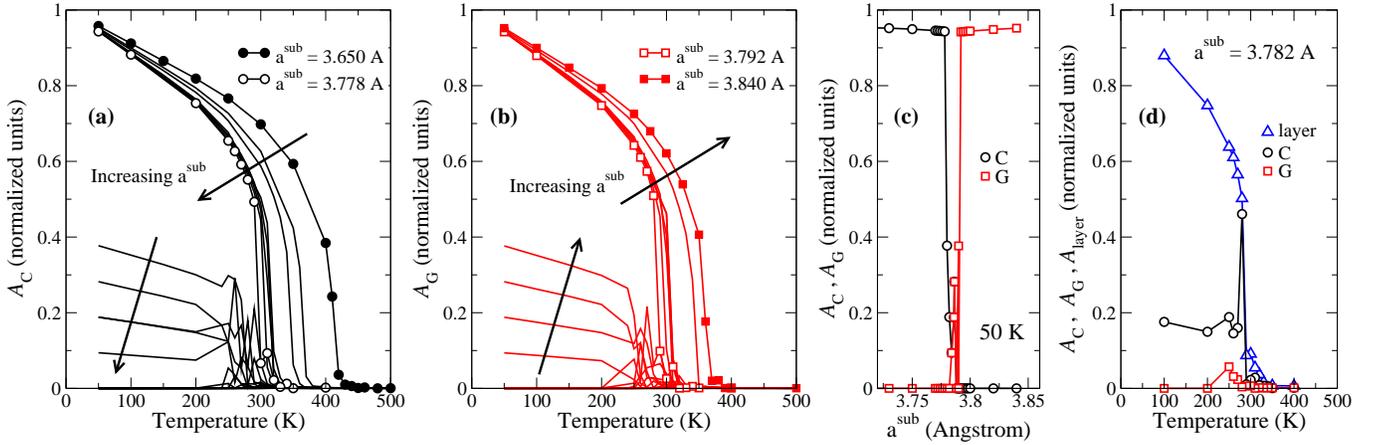}
 \caption{(Color online.) Results from Monte Carlo
   simulations. Panels~(a) and (b): $T$-dependence of the order
   parameters quantifying the degree of C-AFM ($A_{\rm C}$) and G-AFM
   ($A_{\rm G}$) order, respectively. We show the results for all the
   substrates investigated, with $a^{\rm sub}$ ranging from
   3.65~\AA\ to 3.84~\AA. $A_{s} = N^{-1}\sum_{j} S_{jz}
   \exp{(i\boldsymbol{q}_{s} \boldsymbol{R}_{j})}$, where $N$ is the
   number of cells in the simulation box and $\boldsymbol{R}_{j}$ is
   the lattice vector corresponding to spin $\boldsymbol{S}_{j}$;
   $\boldsymbol{q}_s$ defines the $s$-like spin arrangement, with
   $\boldsymbol{q}_{\rm C} = 2\pi/a^{\rm sub} (1/2,1/2,0)$ and
   $\boldsymbol{q}_{\rm G} = 2\pi (1/2a^{\rm sub},1/2a^{\rm
     sub},1/2c)$. We only need to consider the $z$ component of the
   spins ($S_{jz}$) because of the small symmetry-breaking included in
   our Hamiltonians~\protect\cite{anisotropy}. A perfect $s$-like
   order corresponds to having $A_{s}=1$. Panel~(c): $A_{\rm C}$ and
   $A_{\rm G}$ obtained at 50~K and as a function of $a^{\rm
     sub}$. Panel~(d): results for $a^{\rm sub} =$~3.782~\AA. $A_{\rm
     layer}$ quantifies the AFM order within the $ab$ layers. $A_{\rm
     layer} = N_{\rm layer}^{-1}\sum_{j}^{'} S_{jz}
   \exp{(i\boldsymbol{q}_{\rm 2D} \boldsymbol{R}_{j})}$, where the
   primed sum runs over the spins in the first $ab$ layer, which is
   representative of the rest; $N_{\rm layer}$ is the number of cells
   in a layer and $\boldsymbol{q}_{\rm 2D} = 2\pi/a^{\rm sub}
   (1/2,1/2,0)$.}
\end{figure*}

The obtained $a^{\rm sub}$-dependence of the exchange constants
$J_{\rm a}$ and $J_{\rm c}$ seems rather natural. As we compress
in-plane, the Fe spins coupled by $J_{\rm a}$ get closer and their
interaction becomes stronger; in contrast, the distance between irons
coupled by $J_{\rm c}$ grows (results in~\cite{suppl}) and the
interaction weakens~\cite{simplistic}. However, it is not clear
what to expect for $J_{\rm ac}$. In this case, any coupling mechanism
that one can imagine will have both in-plane and out-of-plane {\sl
  components}; hence, it is probably not surprising to find that
$J_{\rm ac}$ varies weakly with $a^{\rm sub}$.

Hence, our calculations and model analysis reveal a robust mechanism
leading to a crossover between G-AFM and C-AFM orders as $a^{\rm sub}$
varies. Such a transition is the result of the competition between two
magnetic interactions that become comparable in a certain range of
epitaxial strains. In fact, our $ST$-BFO films can be considered a
case of {\em frustrated spin system}, where the magnitude of the
frustration can be {\em tuned} by appropriately choosing the substrates
on which the films are grown.

{\sl Phase diagram.--} We solved our Hamiltonians by performing Monte
Carlo simulations in a periodically-repeated box of
20$\times$20$\times$20
spins~\cite{montecarlo,rubtsov00}. Figures~3(a) and 3(b) show,
respectively, our results for the $T$-dependence of the parameters
monitoring the C-AFM ($A_{\rm C}$) and G-AFM ($A_{\rm G}$) orders;
Fig.~3(c) shows the results for $T=$~50~K, where the evolution of the
ground state with $a^{\rm sub}$ is easily
appreciated~\cite{anisotropy}.

Our simulations render a paramagnetic (PM) phase at high
temperatures. Then, for $a^{\rm sub}\leq$~3.778~\AA\ we observe a
transition to a C-AFM phase as $T$ decreases; in contrast, for $a^{\rm
  sub}\geq$~3.792~\AA\ the low-$T$ phase presents the G-AFM spin
order. In the narrow intermediate region
3.778~\AA~$<a^{sub}<$~3.792~\AA, some sort of $T$-driven transition
occurs as evidenced by the non-zero values of $A_{\rm G}$ and $A_{\rm
  C}$; however, no clear-cut three-dimensional order emerges.

Figure~3(d) shows results for $a^{\rm sub}$~=~3.782~\AA, which is
representative of the intermediate region. In this case, $A_{\rm G}$
and $A_{\rm C}$ take values that look rather arbitrary. Nevertheless,
if we consider the AFM order within the $ab$ planes, as quantified by
$A_{\rm layer}$ defined in the figure caption, we recover a
well-behaved transition. The picture that emerges is thus clear: For
intermediate values of $a^{\rm sub}$ the system becomes {\em
  two-dimensional} (2D). The out-of-plane correlations are very weak,
and long-range order along $z$ essentially disappears.

Figure~4 shows the phase diagram that emerges from our results. Note
that the width of the 2D region depends on the specific box size and
sampling method used in our simulations. Indeed, for small but
non-zero values of $|J_{c}-4J_{ac}|$, the ground state of the system
is well defined in the thermodynamic limit. However, the specifics of
our simulations will determine whether the AFM planes can order {\em
  correctly} along the $z$ direction or, instead, get stuck in a
meta-stable configuration displaying disorder and/or phase
co-existence. [As appreciated in Fig.~3(c), we tend to find incomplete
  G-like (resp. C-like) order on the right (left) side of the
  intermediate region, which is a result of having
  $J_{c}-4J_{ac}~\gtrsim$~0 (resp. $J_{c}-4J_{ac}~\lesssim$~0).]
Hence, the 2D region in Fig.~4 has to be taken as evidence for what
probably is a {\em line} separating the C-AFM and G-AFM phases in the
ideal case.

\begin{figure}[t!]
 \includegraphics[width=0.85\columnwidth]{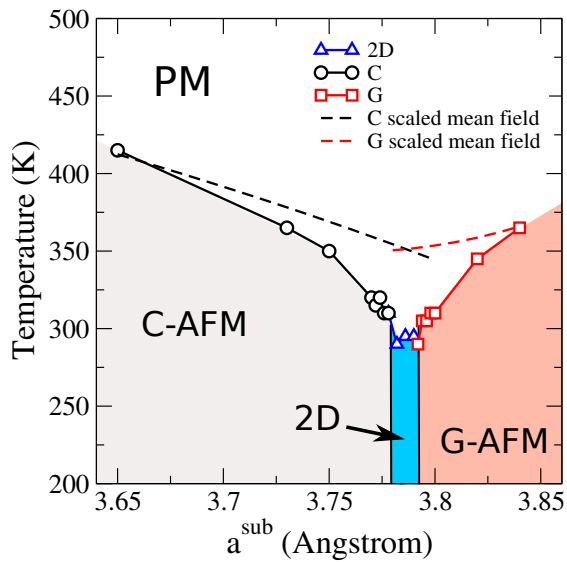}
 \caption{(Color online.) Phase diagram deduced from our Monte Carlo
   simulation. The dashed lines show the transition lines obtained at
   the mean-field level, which are rescaled to fit with the Monte
   Carlo results (see text).}
\end{figure}

In order to understand the evolution of $T_{\rm N}$ with $a^{\rm
  sub}$, it is useful to resort to a mean-field (MF) analysis of our
Hamiltonians. Within this approximation, the transition temperature is
proportional to the {\em mean field} experienced by the spins in the
lattice; such a field is essentially given by $4J_{a}-2J_{c}+8J_{ac}$
in the C-AFM case, and by $4J_{a}+2J_{c}-8J_{ac}$ for the G-AFM
solution. In Fig.~4 we plot the $a^{\rm sub}$-dependence of these MFs;
note that we multiply the results by a factor that is constant through
the whole $a^{\rm sub}$ range and common to both spin arrangements, so
that the MF lines fit in the scale of the Monte Carlos results. As we
can see, the general dependence of $T_{\rm N}$ with $a^{\rm sub}$ is
captured at the MF level. In particular, for decreasing $a^{\rm sub}$
values, $T_{\rm N}$ of the PM$\rightarrow$C-AFM transition grows as a
consequence of the increasing in-plane coupling $J_{a}$. On the other
hand, if $a^{\rm sub}$ becomes larger, $T_{\rm N}$ of the
PM$\rightarrow$G-AFM transition grows driven by the rapidly increasing
$J_{c}$ interaction. Finally, as we approach the 2D region, the
$T_{\rm N}$ values obtained from Monte Carlo decrease more strongly
than we would predict at the MF level. This is most likely a
consequence of the competition at play in the Monte Carlo simulations,
as such an effect is not captured by our simple MF analysis.

{\sl Further discussion.--} Previous works have shown that the
approach adopted here renders reliable N\`eel temperatures for
$ST$-BFO~\cite{dieguez11b,infante11}. Indeed, for BFO films grown on
LaAlO$_3$ ($a^{\rm sub}$~=~3.79~\AA), measured $T_{\rm N}$ values
range from 324~K~\cite{macdougall12} to 360~K~\cite{infante11}, which
is in reasonable agreement with our results~\cite{accuracy,hong12}.

Regarding the specific magnetic order, as far as we know only films
grown on LaAlO$_3$ have been characterized experimentally. There seems
to be consensus about the fact that the films adopt a G-AFM order at
$T_{\rm r}$~\cite{bea09,infante11,macdougall12}. Additionally,
MacDougall {\sl et al}.~\cite{macdougall12} have reported a
coexistence of G-AFM and C-AFM orders at temperatures below
260~K. Such a situation seems compatible with the 2D region that we
find precisely for $a^{\rm sub}\approx$~3.79~\AA. Indeed, for
$J_{c}-4J_{ac}~\gtrsim$~0 we obtained a dominant, but not complete,
G-AFM order [see Fig.~3(c)], which clearly resembles the experimental
findings of Ref.~\onlinecite{macdougall12}.

At and around the 2D region we have $J_{c}-4J_{ac}\approx$~0, which
results in an effective magnetic {\em decoupling} of the $ab$
planes. In such cases, it is conceivable that impurities, defects, and
other extrinsic factors may influence the nature of the magnetic
ground state or the relative populations of co-existing
phases. However, for smaller $a^{\rm sub}$ values we obtain a sizable
$J_{c}-4J_{ac} <$~0. Hence, we predict that C-AFM order will occur in
$SF$-BFO grown on substrates more compressive than LaAlO$_{3}$.

Note that this picture differs significantly from that of MacDougall
{\sl et al}.~\cite{macdougall12}, who argued that all the out-of-plane
couplings are essentially negligible (i.e., $J_{c}\approx$~0 and
$J_{ac}\approx$~0) in $ST$-BFO grown on LaAlO$_{3}$. Their
interpretation implies that the magnetic order in $ST$-BFO should be
2D-like for any substrate more compressive than LaAlO$_3$, and one
would not expect to observe any robust C-AFM phase. This is clearly at
variance with our results.

How general is the tunable geometric frustration predicted in this
work? Let us emphasize that the requirements to observe these effects
do not seem exotic at all: In essence, we need to (1) have AFM
interactions between neighboring magnetic cations, and (2) obtain an
expansion of the out-of-plane lattice parameter as we compress
in-plane. Condition~(1) is satisfied by many perovskite oxides that
are AFM insulators. Condition~(2) is the expected behavior for all
perovskites; further, there is a growing number of compounds that are
known to display large $c/a$ ratios when grown under appropriate
conditions.

To test this presumed generality, we investigated BiCoO$_{3}$ (BCO)
under epitaxial strain. BCO is an insulating perovskite whose ground
state has the ideal $ST$ structure ($P4mm$ space group, with $c/a
=$~1.27 and $a =$3.73~\AA) and a C-AFM spin order~\cite{belik06}. As
we imposed a tensile epitaxial strain, we observed a transition from
C-AFM to G-AFM at $a^{\rm sub}\approx$~3.84~\AA. As in the case of
$SF$-BFO, the analogous crossover between the FM and A-AFM orders is
absent, and the transition point is defined by the condition $J_{c} =
4J_{ac}$, with all the computed exchange couplings being AFM in
nature. Hence, BCO presents exactly the same magnetic frustration
effects that we have discussed for $ST$-BFO. Our BCO results are
summarized in~\cite{suppl}, including the prediction that G-AFM order
can be obtained in BCO films at $T_{\rm r}$. Hence, BCO, as well as
the BFO-BCO solid solutions studied in Ref.~\onlinecite{dieguez11b},
may offer interesting alternatives for the experimental investigation
of these effects.

Work supported by MINECO-Spain through Grants No. MAT2010-18113,
No. MAT2010-10093-E, and No. CSD2007-00041, and the ``Ram\'on y
Cajal'' program (OD). We made use of the computing facilities provided
by CESGA and RES. M.~Bibes' comments are gratefully acknowledged.


\begin{thebibliography}{23}%
\makeatletter
\providecommand \@ifxundefined [1]{%
 \@ifx{#1\undefined}
}%
\providecommand \@ifnum [1]{%
 \ifnum #1\expandafter \@firstoftwo
 \else \expandafter \@secondoftwo
 \fi
}%
\providecommand \@ifx [1]{%
 \ifx #1\expandafter \@firstoftwo
 \else \expandafter \@secondoftwo
 \fi
}%
\providecommand \natexlab [1]{#1}%
\providecommand \enquote  [1]{``#1''}%
\providecommand \bibnamefont  [1]{#1}%
\providecommand \bibfnamefont [1]{#1}%
\providecommand \citenamefont [1]{#1}%
\providecommand \href@noop [0]{\@secondoftwo}%
\providecommand \href [0]{\begingroup \@sanitize@url \@href}%
\providecommand \@href[1]{\@@startlink{#1}\@@href}%
\providecommand \@@href[1]{\endgroup#1\@@endlink}%
\providecommand \@sanitize@url [0]{\catcode `\\12\catcode `\$12\catcode
  `\&12\catcode `\#12\catcode `\^12\catcode `\_12\catcode `\%12\relax}%
\providecommand \@@startlink[1]{}%
\providecommand \@@endlink[0]{}%
\providecommand \url  [0]{\begingroup\@sanitize@url \@url }%
\providecommand \@url [1]{\endgroup\@href {#1}{\urlprefix }}%
\providecommand \urlprefix  [0]{URL }%
\providecommand \Eprint [0]{\href }%
\providecommand \doibase [0]{http://dx.doi.org/}%
\providecommand \selectlanguage [0]{\@gobble}%
\providecommand \bibinfo  [0]{\@secondoftwo}%
\providecommand \bibfield  [0]{\@secondoftwo}%
\providecommand \translation [1]{[#1]}%
\providecommand \BibitemOpen [0]{}%
\providecommand \bibitemStop [0]{}%
\providecommand \bibitemNoStop [0]{.\EOS\space}%
\providecommand \EOS [0]{\spacefactor3000\relax}%
\providecommand \BibitemShut  [1]{\csname bibitem#1\endcsname}%
\let\auto@bib@innerbib\@empty
\bibitem [{\citenamefont {Tokura}\ and\ \citenamefont
  {Tomioka}(1999)}]{tokura99}%
  \BibitemOpen
  \bibfield  {author} {\bibinfo {author} {\bibfnamefont {Y.}~\bibnamefont
  {Tokura}}\ and\ \bibinfo {author} {\bibfnamefont {Y.}~\bibnamefont
  {Tomioka}},\ }\href {\doibase 10.1016/S0304-8853(99)00352-2} {\bibfield
  {journal} {\bibinfo  {journal} {Journal of Magnetism and Magnetic Materials}\
  }\textbf {\bibinfo {volume} {200}},\ \bibinfo {pages} {1} (\bibinfo {year}
  {1999})}\BibitemShut {NoStop}%
\bibitem [{\citenamefont {Pirc}\ and\ \citenamefont {Blinc}(1999)}]{pirc99}%
  \BibitemOpen
  \bibfield  {author} {\bibinfo {author} {\bibfnamefont {R.}~\bibnamefont
  {Pirc}}\ and\ \bibinfo {author} {\bibfnamefont {R.}~\bibnamefont {Blinc}},\
  }\href {\doibase 10.1103/PhysRevB.60.13470} {\bibfield  {journal} {\bibinfo
  {journal} {Physical Review B}\ }\textbf {\bibinfo {volume} {60}},\ \bibinfo
  {pages} {13470} (\bibinfo {year} {1999})}\BibitemShut {NoStop}%
\bibitem [{\citenamefont {Samara}(2003)}]{samara03}%
  \BibitemOpen
  \bibfield  {author} {\bibinfo {author} {\bibfnamefont {G.~A.}\ \bibnamefont
  {Samara}},\ }\href {\doibase 10.1088/0953-8984/15/9/202} {\bibfield
  {journal} {\bibinfo  {journal} {Journal of Physics: Condensed Matter}\
  }\textbf {\bibinfo {volume} {15}},\ \bibinfo {pages} {R367} (\bibinfo {year}
  {2003})}\BibitemShut {NoStop}%
\bibitem [{\citenamefont {Catalan}\ and\ \citenamefont
  {Scott}(2009)}]{catalan09}%
  \BibitemOpen
  \bibfield  {author} {\bibinfo {author} {\bibfnamefont {G.}~\bibnamefont
  {Catalan}}\ and\ \bibinfo {author} {\bibfnamefont {J.~F.}\ \bibnamefont
  {Scott}},\ }\href {\doibase 10.1002/adma.200802849} {\bibfield  {journal}
  {\bibinfo  {journal} {Advanced Materials}\ }\textbf {\bibinfo {volume}
  {21}},\ \bibinfo {pages} {2463} (\bibinfo {year} {2009})}\BibitemShut
  {NoStop}%
\bibitem [{\citenamefont {{B\'ea}}\ \emph {et~al.}(2009)\citenamefont
  {{B\'ea}}, \citenamefont {Dupe}, \citenamefont {Fusil}, \citenamefont
  {Mattana}, \citenamefont {Jacquet}, \citenamefont {{Warot-Fonrose}},
  \citenamefont {Wilhelm}, \citenamefont {Rogalev}, \citenamefont {Petit},
  \citenamefont {Cros}, \citenamefont {Anane}, \citenamefont {Petroff},
  \citenamefont {Bouzehouane}, \citenamefont {Geneste}, \citenamefont {Dkhil},
  \citenamefont {Lisenkov}, \citenamefont {Ponomareva}, \citenamefont
  {Bellaiche}, \citenamefont {Bibes},\ and\ \citenamefont
  {{Barth\'el\'emy}}}]{bea09}%
  \BibitemOpen
  \bibfield  {author} {\bibinfo {author} {\bibfnamefont {H.}~\bibnamefont
  {{B\'ea}}}, \bibinfo {author} {\bibfnamefont {B.}~\bibnamefont {Dupe}},
  \bibinfo {author} {\bibfnamefont {S.}~\bibnamefont {Fusil}}, \bibinfo
  {author} {\bibfnamefont {R.}~\bibnamefont {Mattana}}, \bibinfo {author}
  {\bibfnamefont {E.}~\bibnamefont {Jacquet}}, \bibinfo {author} {\bibfnamefont
  {B.}~\bibnamefont {{Warot-Fonrose}}}, \bibinfo {author} {\bibfnamefont
  {F.}~\bibnamefont {Wilhelm}}, \bibinfo {author} {\bibfnamefont
  {A.}~\bibnamefont {Rogalev}}, \bibinfo {author} {\bibfnamefont
  {S.}~\bibnamefont {Petit}}, \bibinfo {author} {\bibfnamefont
  {V.}~\bibnamefont {Cros}}, \bibinfo {author} {\bibfnamefont {A.}~\bibnamefont
  {Anane}}, \bibinfo {author} {\bibfnamefont {F.}~\bibnamefont {Petroff}},
  \bibinfo {author} {\bibfnamefont {K.}~\bibnamefont {Bouzehouane}}, \bibinfo
  {author} {\bibfnamefont {G.}~\bibnamefont {Geneste}}, \bibinfo {author}
  {\bibfnamefont {B.}~\bibnamefont {Dkhil}}, \bibinfo {author} {\bibfnamefont
  {S.}~\bibnamefont {Lisenkov}}, \bibinfo {author} {\bibfnamefont
  {I.}~\bibnamefont {Ponomareva}}, \bibinfo {author} {\bibfnamefont
  {L.}~\bibnamefont {Bellaiche}}, \bibinfo {author} {\bibfnamefont
  {M.}~\bibnamefont {Bibes}}, \ and\ \bibinfo {author} {\bibfnamefont
  {A.}~\bibnamefont {{Barth\'el\'emy}}},\ }\href@noop {} {\bibfield  {journal}
  {\bibinfo  {journal} {Physical Review Letters}\ }\textbf {\bibinfo {volume}
  {102}},\ \bibinfo {pages} {217603} (\bibinfo {year} {2009})}\BibitemShut
  {NoStop}%
\bibitem [{\citenamefont {Infante}\ \emph {et~al.}(2011)\citenamefont
  {Infante}, \citenamefont {Juraszek}, \citenamefont {Fusil}, \citenamefont
  {{Dup\'e}}, \citenamefont {Gemeiner}, \citenamefont {{Di\'eguez}},
  \citenamefont {Pailloux}, \citenamefont {Jouen}, \citenamefont {Jacquet},
  \citenamefont {Geneste}, \citenamefont {Pacaud}, \citenamefont
  {{\'I\~niguez}}, \citenamefont {Bellaiche}, \citenamefont {{Barth\'el\'emy}},
  \citenamefont {Dkhil},\ and\ \citenamefont {Bibes}}]{infante11}%
  \BibitemOpen
  \bibfield  {author} {\bibinfo {author} {\bibfnamefont {I.~C.}\ \bibnamefont
  {Infante}}, \bibinfo {author} {\bibfnamefont {J.}~\bibnamefont {Juraszek}},
  \bibinfo {author} {\bibfnamefont {S.}~\bibnamefont {Fusil}}, \bibinfo
  {author} {\bibfnamefont {B.}~\bibnamefont {{Dup\'e}}}, \bibinfo {author}
  {\bibfnamefont {P.}~\bibnamefont {Gemeiner}}, \bibinfo {author}
  {\bibfnamefont {O.}~\bibnamefont {{Di\'eguez}}}, \bibinfo {author}
  {\bibfnamefont {F.}~\bibnamefont {Pailloux}}, \bibinfo {author}
  {\bibfnamefont {S.}~\bibnamefont {Jouen}}, \bibinfo {author} {\bibfnamefont
  {E.}~\bibnamefont {Jacquet}}, \bibinfo {author} {\bibfnamefont
  {G.}~\bibnamefont {Geneste}}, \bibinfo {author} {\bibfnamefont
  {J.}~\bibnamefont {Pacaud}}, \bibinfo {author} {\bibfnamefont
  {J.}~\bibnamefont {{\'I\~niguez}}}, \bibinfo {author} {\bibfnamefont
  {L.}~\bibnamefont {Bellaiche}}, \bibinfo {author} {\bibfnamefont
  {A.}~\bibnamefont {{Barth\'el\'emy}}}, \bibinfo {author} {\bibfnamefont
  {B.}~\bibnamefont {Dkhil}}, \ and\ \bibinfo {author} {\bibfnamefont
  {M.}~\bibnamefont {Bibes}},\ }\href {\doibase 10.1103/PhysRevLett.107.237601}
  {\bibfield  {journal} {\bibinfo  {journal} {Physical Review Letters}\
  }\textbf {\bibinfo {volume} {107}},\ \bibinfo {pages} {237601} (\bibinfo
  {year} {2011})}\BibitemShut {NoStop}%
\bibitem [{\citenamefont {Ko}\ \emph {et~al.}(2011)\citenamefont {Ko},
  \citenamefont {Jung}, \citenamefont {He}, \citenamefont {Lee}, \citenamefont
  {Woo}, \citenamefont {Chu}, \citenamefont {Seidel}, \citenamefont {Jeon},
  \citenamefont {Oh}, \citenamefont {Kim}, \citenamefont {Liang}, \citenamefont
  {Chen}, \citenamefont {Chu}, \citenamefont {Jeong}, \citenamefont {Ramesh},
  \citenamefont {Park},\ and\ \citenamefont {Yang}}]{ko11}%
  \BibitemOpen
  \bibfield  {author} {\bibinfo {author} {\bibfnamefont {K.}~\bibnamefont
  {Ko}}, \bibinfo {author} {\bibfnamefont {M.~H.}\ \bibnamefont {Jung}},
  \bibinfo {author} {\bibfnamefont {Q.}~\bibnamefont {He}}, \bibinfo {author}
  {\bibfnamefont {J.~H.}\ \bibnamefont {Lee}}, \bibinfo {author} {\bibfnamefont
  {C.~S.}\ \bibnamefont {Woo}}, \bibinfo {author} {\bibfnamefont
  {K.}~\bibnamefont {Chu}}, \bibinfo {author} {\bibfnamefont {J.}~\bibnamefont
  {Seidel}}, \bibinfo {author} {\bibfnamefont {B.}~\bibnamefont {Jeon}},
  \bibinfo {author} {\bibfnamefont {Y.~S.}\ \bibnamefont {Oh}}, \bibinfo
  {author} {\bibfnamefont {K.~H.}\ \bibnamefont {Kim}}, \bibinfo {author}
  {\bibfnamefont {W.}~\bibnamefont {Liang}}, \bibinfo {author} {\bibfnamefont
  {H.}~\bibnamefont {Chen}}, \bibinfo {author} {\bibfnamefont {Y.}~\bibnamefont
  {Chu}}, \bibinfo {author} {\bibfnamefont {Y.~H.}\ \bibnamefont {Jeong}},
  \bibinfo {author} {\bibfnamefont {R.}~\bibnamefont {Ramesh}}, \bibinfo
  {author} {\bibfnamefont {J.}~\bibnamefont {Park}}, \ and\ \bibinfo {author}
  {\bibfnamefont {C.}~\bibnamefont {Yang}},\ }\href {\doibase
  10.1038/ncomms1576} {\bibfield  {journal} {\bibinfo  {journal} {Nature
  Communications}\ }\textbf {\bibinfo {volume} {2}},\ \bibinfo {pages} {567}
  (\bibinfo {year} {2011})}\BibitemShut {NoStop}%
\bibitem [{\citenamefont {Hatt}\ \emph {et~al.}(2010)\citenamefont {Hatt},
  \citenamefont {Spaldin},\ and\ \citenamefont {Ederer}}]{hatt10}%
  \BibitemOpen
  \bibfield  {author} {\bibinfo {author} {\bibfnamefont {A.~J.}\ \bibnamefont
  {Hatt}}, \bibinfo {author} {\bibfnamefont {N.~A.}\ \bibnamefont {Spaldin}}, \
  and\ \bibinfo {author} {\bibfnamefont {C.}~\bibnamefont {Ederer}},\ }\href
  {\doibase 10.1103/PhysRevB.81.054109} {\bibfield  {journal} {\bibinfo
  {journal} {Physical Review B}\ }\textbf {\bibinfo {volume} {81}},\ \bibinfo
  {pages} {054109} (\bibinfo {year} {2010})}\BibitemShut {NoStop}%
\bibitem [{\citenamefont {{Di\'eguez}}\ \emph {et~al.}(2011)\citenamefont
  {{Di\'eguez}}, \citenamefont {{Gonz\'alez-V\'azquez}}, \citenamefont
  {{Wojde\l}},\ and\ \citenamefont {{\'I\~niguez}}}]{dieguez11}%
  \BibitemOpen
  \bibfield  {author} {\bibinfo {author} {\bibfnamefont {O.}~\bibnamefont
  {{Di\'eguez}}}, \bibinfo {author} {\bibfnamefont {O.~E.}\ \bibnamefont
  {{Gonz\'alez-V\'azquez}}}, \bibinfo {author} {\bibfnamefont {J.~C.}\
  \bibnamefont {{Wojde\l}}}, \ and\ \bibinfo {author} {\bibfnamefont
  {J.}~\bibnamefont {{\'I\~niguez}}},\ }\href@noop {} {\bibfield  {journal}
  {\bibinfo  {journal} {Physical Review B}\ }\textbf {\bibinfo {volume} {83}},\
  \bibinfo {pages} {094105} (\bibinfo {year} {2011})}\BibitemShut {NoStop}%
\bibitem [{\citenamefont {{MacDougall}}\ \emph {et~al.}(2012)\citenamefont
  {{MacDougall}}, \citenamefont {Christen}, \citenamefont {Siemons},
  \citenamefont {Biegalski}, \citenamefont {Zarestky}, \citenamefont {Liang},
  \citenamefont {Dagotto},\ and\ \citenamefont {Nagler}}]{macdougall12}%
  \BibitemOpen
  \bibfield  {author} {\bibinfo {author} {\bibfnamefont {G.~J.}\ \bibnamefont
  {{MacDougall}}}, \bibinfo {author} {\bibfnamefont {H.~M.}\ \bibnamefont
  {Christen}}, \bibinfo {author} {\bibfnamefont {W.}~\bibnamefont {Siemons}},
  \bibinfo {author} {\bibfnamefont {M.~D.}\ \bibnamefont {Biegalski}}, \bibinfo
  {author} {\bibfnamefont {J.~L.}\ \bibnamefont {Zarestky}}, \bibinfo {author}
  {\bibfnamefont {S.}~\bibnamefont {Liang}}, \bibinfo {author} {\bibfnamefont
  {E.}~\bibnamefont {Dagotto}}, \ and\ \bibinfo {author} {\bibfnamefont
  {S.~E.}\ \bibnamefont {Nagler}},\ }\href {\doibase
  10.1103/PhysRevB.85.100406} {\bibfield  {journal} {\bibinfo  {journal}
  {Physical Review B}\ }\textbf {\bibinfo {volume} {85}},\ \bibinfo {pages}
  {100406} (\bibinfo {year} {2012})}\BibitemShut {NoStop}%
\bibitem [{\citenamefont {{Di\'eguez}}\ and\ \citenamefont
  {{\'I\~niguez}}(2011)}]{dieguez11b}%
  \BibitemOpen
  \bibfield  {author} {\bibinfo {author} {\bibfnamefont {O.}~\bibnamefont
  {{Di\'eguez}}}\ and\ \bibinfo {author} {\bibfnamefont {J.}~\bibnamefont
  {{\'I\~niguez}}},\ }\href {\doibase 10.1103/PhysRevLett.107.057601}
  {\bibfield  {journal} {\bibinfo  {journal} {Physical Review Letters}\
  }\textbf {\bibinfo {volume} {107}},\ \bibinfo {pages} {057601} (\bibinfo
  {year} {2011})}\BibitemShut {NoStop}%
\bibitem [{cal()}]{calcs}%
  \BibitemOpen
  \href@noop {} {}\bibinfo {note} {The first-principles methods employed are
  described in the online Supplementary Materials~\protect\cite{suppl}, and are
  exactly the ones used in Ref.~\protect\onlinecite{dieguez11b}.}\BibitemShut
  {Stop}%
\bibitem [{sup()}]{suppl}%
  \BibitemOpen
  \href@noop {} {}\bibinfo {note} {See online Supplementary Materials
  accompanying this article.}\BibitemShut {Stop}%
\bibitem [{\citenamefont {Christen}\ \emph {et~al.}(2011)\citenamefont
  {Christen}, \citenamefont {Nam}, \citenamefont {Kim}, \citenamefont {Hatt},\
  and\ \citenamefont {Spaldin}}]{christen11}%
  \BibitemOpen
  \bibfield  {author} {\bibinfo {author} {\bibfnamefont {H.~M.}\ \bibnamefont
  {Christen}}, \bibinfo {author} {\bibfnamefont {J.~H.}\ \bibnamefont {Nam}},
  \bibinfo {author} {\bibfnamefont {H.~S.}\ \bibnamefont {Kim}}, \bibinfo
  {author} {\bibfnamefont {A.~J.}\ \bibnamefont {Hatt}}, \ and\ \bibinfo
  {author} {\bibfnamefont {N.~A.}\ \bibnamefont {Spaldin}},\ }\href {\doibase
  10.1103/PhysRevB.83.144107} {\bibfield  {journal} {\bibinfo  {journal}
  {Physical Review B}\ }\textbf {\bibinfo {volume} {83}},\ \bibinfo {pages}
  {144107} (\bibinfo {year} {2011})}\BibitemShut {NoStop}%
\bibitem [{asu()}]{asub}%
  \BibitemOpen
  \href@noop {} {}\bibinfo {note} {We ran first-principles calculations for
  $a^{\rm sub}$ values in a grid with $\Delta a^{\rm sub}$~=~0.01~\AA. To
  determine the boundaries of the 2D region of Fig.~4, we ran Monte Carlo
  simulations for an even finer grid with $\Delta a^{\rm sub}$~=~0.002~\AA; the
  employed spin Hamiltonians were obtained by interpolating the results in
  Fig.~2(b).}\BibitemShut {Stop}%
\bibitem [{\citenamefont {Khomskii}(2001)}]{khomskii01}%
  \BibitemOpen
  \bibfield  {author} {\bibinfo {author} {\bibfnamefont {D.}~\bibnamefont
  {Khomskii}},\ }in\ \href@noop {} {\emph {\bibinfo {booktitle} {Spin
  Electronics}}},\ \bibinfo {series} {Lecture Notes in Physics}, Vol.\ \bibinfo
  {volume} {569},\ \bibinfo {editor} {edited by\ \bibinfo {editor}
  {\bibfnamefont {M.}~\bibnamefont {Ziese}}\ and\ \bibinfo {editor}
  {\bibfnamefont {M.}~\bibnamefont {Thornton}}}\ (\bibinfo  {publisher}
  {Springer Berlin / Heidelberg},\ \bibinfo {year} {2001})\ pp.\ \bibinfo
  {pages} {89--116}\BibitemShut {NoStop}%
\bibitem [{ani()}]{anisotropy}%
  \BibitemOpen
  \href@noop {} {}\bibinfo {note} {For the sake of convenience, we introduced a
  small magnetic anisotropy in the spin Hamiltonians, so that the easy axis is
  fixed to lie along $z$. Thus, in Fig.~3 we only show the $z$ component of the
  three-dimensional order parameters.}\BibitemShut {Stop}%
\bibitem [{sim()}]{simplistic}%
  \BibitemOpen
  \href@noop {} {}\bibinfo {note} {The qualitative $a^{\rm sub}$-dependence of
  $J_{\rm a}$ and $J_{\rm c}$ clearly correlates with the variation of the
  corresponding the Fe--Fe distances. This trivial dependence was also pointed
  out by MacDougall {\sl et al}.~\protect\cite{macdougall12} when discussing
  the differences in the magnetic couplings between BFO's rhombohedral and $ST$
  phases. Of course, a more quantitative argument should take into account the
  detailed evolution of the interaction paths (e.g., the $a^{\rm
  sub}$-dependence of the Fe--O--Fe angle in the case of $J_{\rm a}$, and of
  the two different Fe--O distances in the case of $J_{\rm c}$.)}\BibitemShut
  {NoStop}%
\bibitem [{mon()}]{montecarlo}%
  \BibitemOpen
  \href@noop {} {}\bibinfo {note} {We ran regular Metropolis Monte Carlo,
  performing 10000~sweeps of the full simulation box for thermalization and
  50000 additional sweeps for computing averages. The algorithm to generate new
  spin configurations was a combination of regular sampling (typically aiming
  at an acceptance ratio of 40~\%) and the {\em magic steps} described in
  Ref.~\protect\onlinecite{rubtsov00}. The calculation conditions were checked
  to render sufficiently converged results.}\BibitemShut {Stop}%
\bibitem [{\citenamefont {Rubtsov}\ \emph {et~al.}(2000)\citenamefont
  {Rubtsov}, \citenamefont {Hlinka},\ and\ \citenamefont
  {Janssen}}]{rubtsov00}%
  \BibitemOpen
  \bibfield  {author} {\bibinfo {author} {\bibfnamefont {A.~N.}\ \bibnamefont
  {Rubtsov}}, \bibinfo {author} {\bibfnamefont {J.}~\bibnamefont {Hlinka}}, \
  and\ \bibinfo {author} {\bibfnamefont {T.}~\bibnamefont {Janssen}},\
  }\href@noop {} {\bibfield  {journal} {\bibinfo  {journal} {Physical Review
  E}\ }\textbf {\bibinfo {volume} {61}},\ \bibinfo {pages} {126} (\bibinfo
  {year} {2000})}\BibitemShut {NoStop}%
\bibitem [{acc()}]{accuracy}%
  \BibitemOpen
  \href@noop {} {}\bibinfo {note} {This level of agreement probably exceeds
  what one may expect from magnetic interactions computed using density
  functional theory (DFT) methods. Yet, note that our DFT scheme, which employs
  a ``Hubbard $U$'' correction with $U = 4$~eV for Fe and $U = 6$~eV for Co,
  was checked to produce magnetic couplings in good agreement with results
  obtained using hybrid functionals, which are known to be accurate for the
  calculation of such interactions in magnetic insulators. The reliability of
  such an $U$-fitting procedure has been recently
  shown~\protect\cite{hong12}.}\BibitemShut {Stop}%
\bibitem [{\citenamefont {Hong}\ \emph {et~al.}(2012)\citenamefont {Hong},
  \citenamefont {Stroppa}, \citenamefont {{\'I\~niguez}}, \citenamefont
  {Picozzi},\ and\ \citenamefont {Vanderbilt}}]{hong12}%
  \BibitemOpen
  \bibfield  {author} {\bibinfo {author} {\bibfnamefont {J.}~\bibnamefont
  {Hong}}, \bibinfo {author} {\bibfnamefont {A.}~\bibnamefont {Stroppa}},
  \bibinfo {author} {\bibfnamefont {J.}~\bibnamefont {{\'I\~niguez}}}, \bibinfo
  {author} {\bibfnamefont {S.}~\bibnamefont {Picozzi}}, \ and\ \bibinfo
  {author} {\bibfnamefont {D.}~\bibnamefont {Vanderbilt}},\ }\href {\doibase
  10.1103/PhysRevB.85.054417} {\bibfield  {journal} {\bibinfo  {journal}
  {Physical Review B}\ }\textbf {\bibinfo {volume} {85}},\ \bibinfo {pages}
  {054417} (\bibinfo {year} {2012})}\BibitemShut {NoStop}%
\bibitem [{\citenamefont {Belik}\ \emph {et~al.}(2006)\citenamefont {Belik},
  \citenamefont {Iikubo}, \citenamefont {Kodama}, \citenamefont {Igawa},
  \citenamefont {Shamoto}, \citenamefont {Niitaka}, \citenamefont {Azuma},
  \citenamefont {Shimakawa}, \citenamefont {Takano}, \citenamefont {Izumi},\
  and\ \citenamefont {Takayama-Muromachi}}]{belik06}%
  \BibitemOpen
  \bibfield  {author} {\bibinfo {author} {\bibfnamefont {A.~A.}\ \bibnamefont
  {Belik}}, \bibinfo {author} {\bibfnamefont {S.}~\bibnamefont {Iikubo}},
  \bibinfo {author} {\bibfnamefont {K.}~\bibnamefont {Kodama}}, \bibinfo
  {author} {\bibfnamefont {N.}~\bibnamefont {Igawa}}, \bibinfo {author}
  {\bibfnamefont {S.}~\bibnamefont {Shamoto}}, \bibinfo {author} {\bibfnamefont
  {S.}~\bibnamefont {Niitaka}}, \bibinfo {author} {\bibfnamefont
  {M.}~\bibnamefont {Azuma}}, \bibinfo {author} {\bibfnamefont
  {Y.}~\bibnamefont {Shimakawa}}, \bibinfo {author} {\bibfnamefont
  {M.}~\bibnamefont {Takano}}, \bibinfo {author} {\bibfnamefont
  {F.}~\bibnamefont {Izumi}}, \ and\ \bibinfo {author} {\bibfnamefont
  {E.}~\bibnamefont {Takayama-Muromachi}},\ }\href {\doibase 10.1021/cm052334z}
  {\bibfield  {journal} {\bibinfo  {journal} {Chemistry of Materials}\ }\textbf
  {\bibinfo {volume} {18}},\ \bibinfo {pages} {798} (\bibinfo {year}
  {2006})}\BibitemShut {NoStop}%
\end{thebibliography}

%

\end{document}